\documentclass[showpacs,oneside,twocolumn,prl,amsmath,amssymb,fleqn]{revtex4-1}
\usepackage{cases}
\usepackage{amsmath}
\usepackage{amssymb}
\usepackage{amsfonts}
\usepackage{amssymb}
\usepackage{dcolumn}
\usepackage{bm}
\usepackage{bbm}
\usepackage{graphicx}
\usepackage{xcolor}
\usepackage{array}
\usepackage{subfigure}
\usepackage[none]{hyphenat}

\begin{document}

\title{Demonstration of Motion Transduction Based on Parametrically Coupled Mechanical Resonators}
\author{Pu Huang}
\affiliation{Hefei National Laboratory for Physics Sciences at
Microscale and Department of Modern Physics, University of Science
and Technology of China, Hefei, 230026, China}
\author{Pengfei Wang}
\affiliation{Hefei National Laboratory for Physics Sciences at
Microscale and Department of Modern Physics, University of Science
and Technology of China, Hefei, 230026, China}
\author{Jingwei Zhou}
\affiliation{Hefei National Laboratory for Physics Sciences at
Microscale and Department of Modern Physics, University of Science
and Technology of China, Hefei, 230026, China}
\author{Zixiang Wang}
\affiliation{Hefei National Laboratory for Physics Sciences at
Microscale and Department of Modern Physics, University of Science
and Technology of China, Hefei, 230026, China}
\author{Chenyong~Ju}
\affiliation{Hefei National Laboratory for Physics Sciences at
Microscale and Department of Modern Physics, University of Science
and Technology of China, Hefei, 230026, China}
\author{Zimeng Wang}
\affiliation{Hefei National Laboratory for Physics Sciences at
Microscale and Department of Modern Physics, University of Science
and Technology of China, Hefei, 230026, China}
\author{Yang Shen}
\affiliation{Hefei National Laboratory for Physics Sciences at
Microscale and Department of Modern Physics, University of Science
and Technology of China, Hefei, 230026, China}
\author{Changkui Duan}
\affiliation{Hefei National Laboratory for Physics Sciences at
Microscale and Department of Modern Physics, University of Science
and Technology of China, Hefei, 230026, China}
\author{Jiangfeng Du}
\altaffiliation{djf@ustc.edu.cn}
\affiliation{Hefei National Laboratory for Physics Sciences at
Microscale and Department of Modern Physics, University of Science
and Technology of China, Hefei, 230026, China}

\begin{abstract}
Universal sensing the motion of mechanical resonators with high
precision and low back-action is of paramount importance in ultra-weak
signal detection which plays a fundamental role in modern physics.
Here we present a universal scheme that transfer mechanically the motion of the resonator not
directly measurable to the one can be precisely measured using mechanical frequency conversion.
Demonstration of the scheme at room temperature shows
that both the motion imprecision and the back-action force are below the intrinsic level
of the objective resonator, which agree well with our theoretical prediction. The scheme developed here
provides an effective interface between an arbitrary mechanical resonator and a high
quantum efficient displacement sensor, and is expected to find extensive applications in high-demanding mechanical-based
force measurements.

\end{abstract}


\maketitle

Mechanical resonators have been used as sensitive detectors of weak force generated from
small charge \cite {Cleland},
single electron spin \cite{Rugar}, the acceleration of a single gold atom \cite {Jensen},
and quantilized magnetic flux \cite {Harris}.
Generally speaking, to reach a higher force sensitivity,
a smaller mechanical resonator is needed.
Mechanical resonators at nano- and subnano- scale
based on carbon nanotube \cite {CNT},
nanowire \cite{Nichol} and multi- and single- layer graphene \cite {Bunch}
have recently been demonstrated as potentially suitable force sensors.
However, scaling down the size of mechanical resonator
makes the detection of its motion a big challenge.

Quantum mechanics tells us that the measurement of position brings about back-action to the mechanical object.
The product of sensitivity and the back-action known as noise product (NP) are used to characterize the measurement process \cite{Clerk}.
Among the state-of-art displacement sensors, the one based on optical photon
has been shown to have very high displacement sensitivity \cite{Arcizet,Verlot}
as well as a controllable NP \cite{cooling,Schliesser}, where the typical
size of the mechanical resonator is much larger than optical wavelength.
Further reduction of the size of the resonator to sub-wavelength level results into the reduction in photon-mechanical coupling proportionally, and so a much stronger optical power is needed to offset the reduction, which could bring additional noise,
such as heating effect, and would further limit the sensitivity \cite{Nichol}.
Non-optical methods have recently been developed to detect the motion of nano-mechanical resonator by
using single-electron transistor \cite {Knobel,LaHaye}, quantum point contact \cite {Cleland1,Poggio},
superconducting interferometer \cite {Etaki}, microwave resonator \cite {Regal}, and single electron spin \cite {Lukin}.
With further improvements, especially those based on optimizing resonator-detector geometry,
not only high sensitivity, but also very low NP has been demonstrated \cite{cooling,jacobs,Teufel,Schliesser}.
If the coupling between these remarkable but objects-dependent detectors with the motion of
arbitrary mechanical resonators is realized efficiently, these schemes will have wide applications.
However, the exploration of such an interface remains elusive.
The solution is to realize an efficient motion transduction between arbitrary mechanical resonators.
In contrast to the coupling to quantum mechanics objects via optical and microwave photons etc.\ \cite{Arcizet,Verlot,Regal,Teufel,Wei2006},
the coupling between mechanical objects via classical forces, such as electrostatic force \cite{Roukes1,Roukes2},
dielectric force \cite{Die1,Die2} and electrothermal force \cite{thermal}, is more universal, in that it has no
stringent limitations on size, geometry and material of the mechanical systems,
and so provides a better choice for motion transduction.

Here we report a motion transduction scheme based on mechanical parametric coupling and its experimental demonstration.
A transduction process with a displacement
transduction imprecision of $3.9 \times 10^{-11}$ m$/\sqrt {\rm Hz}$ and a
back-action force less than $0.8\times10^{-15}$ N$/\sqrt {\rm Hz}$ have been achieved at room temperature.
Both the imprecision and the back-action force are lower than the intrinsic noise
levels of the low frequency resonator being
transferred, ensuring the transduction efficiency.

Our motion transduction device consists of two parametrically coupled mechanical resonators: a target resonator T, whose motion is being transduced, and a detector resonator D, whose response in terms of force is the destination of the transduction. The dynamics of the system can be described classically by the equations for the displacements $x_{\rm t}$ and $x_{\rm d}$ for resonators T and D, respectively, as follows:
\begin{equation}
\begin{split}
\label{eom_cl}
&m_{\rm t} \ddot{x}_{\rm t}+ \Gamma_{\rm t} \dot{x}_{\rm t} + k_{\rm t} x_{\rm t} = -\eta(t) (x_{\rm t} - x_{\rm d} )\\
&m_{\rm d} \ddot{x}_{\rm d}+ \Gamma_{\rm d} \dot{x}_{\rm d} + k_{\rm d} x_{\rm d} = -\eta(t) (x_{\rm d} - x_{\rm t} ).
\end{split}
\end{equation}
Here $m_{\rm t,d}$, $k_{\rm t,d}$ and $\Gamma_{\rm t,d}$
are resonators T and D's effective masses, spring constants, and intrinsic
dissipation rates, respectively. $\eta(t) = 2\eta \cos(\omega_{\rm pu}t)$ is the pumping field to realize parametrical coupling between the two resonators with
$\eta$ the coupling strength
 and $\omega_{\rm pu}$ the pumping frequency  . $\Gamma_{\rm t,d}$
 is related to the resonators' natural (angular) frequencies $\omega_{\rm t,d}$ and quality factors $Q_{\rm t,d}$ by
  $\Gamma_{\rm t,d} = m_{\rm t,d} \omega_{\rm t,d}/Q_{\rm t,d}$. The case of interest here is $\omega_{\rm d} > \omega_{\rm t}$,
  with the frequency of pumping field $\omega_{\rm pu}$ satisfying the frequency conversion (denoted as FC hereafter) condition:
\begin{equation}
\label{resc}
\omega_{\rm pu}=\omega_{\rm d}-\omega_{\rm t}.
\end{equation}
The motion excited on the resonator T is up-converted to the motion of resonator D for detection. FC has been studied in various
physical system such as nonlinear optics and recently been realized in micro-mechanical system \cite{mixer}.
The noise level of the transduction process is characterized by two parameters, the transduction
imprecision $\sqrt{S_{x}^{\rm im}}$ which is equivalent to the measurement imprecision if resonator D is taken as a
force-based displacement sensor, and the back-action force $\sqrt{S_{F}^{\rm ba}}$. A calculation using the input-output theory (see Supplementary information for details) leads to\cite{note}
\begin{equation}
\label{trans}
S_{F}^{\rm ba}S_{x}^{\rm im}=S_{x,\rm d}^{\rm th}S_{F,\rm d}^{\rm th}=4\hbar^{2}(n_{\rm d}+1)^2,
\end{equation}
where $S^{\rm th}_{x, \rm d}$, $S^{\rm th}_{F,\rm d}$, and $n_{\rm d}$ are the resonator D's intrinsic thermal noise densities of displacement and force and the phonon number, respectively.  At a temperature low enough, Eq.\ (\ref{trans}) leads to the quantum mechanical limit, as expected. A product similar to the second equality of Eq.\ (\ref{trans}) holds for resonator T. At high temperatures, $n_{t,d} \approx k_{\rm B}T/(\hbar \omega_{t,d}) \gg 1$,
we obtain using Eq.\ (\ref{trans}) $\sqrt{{S_{x}^{\rm im}S_{F}^{\rm ba}}/({S_{x,\rm t}^{\rm th}S_{F,\rm t}^{\rm th}})} \propto {\omega_{\rm t}}/{\omega_{\rm d}}$.
Here $S_{x, \rm t}^{\rm th} $ and $S_{F,\rm t}^{\rm th} $ are resonator T's intrinsic thermal noise densities of the displacement and force, respectively. This shows that a resonator D with higher frequency is the actually requirement for high transduction efficiency. For the case of weak force sensing, although the dissipation could set indirectly some constraints on the required effective coupling strength $\eta$, we show that physical realization is feasible (see Supplementary information for details).
\begin{figure}[htbp]
\centering
\includegraphics[width=1\columnwidth]{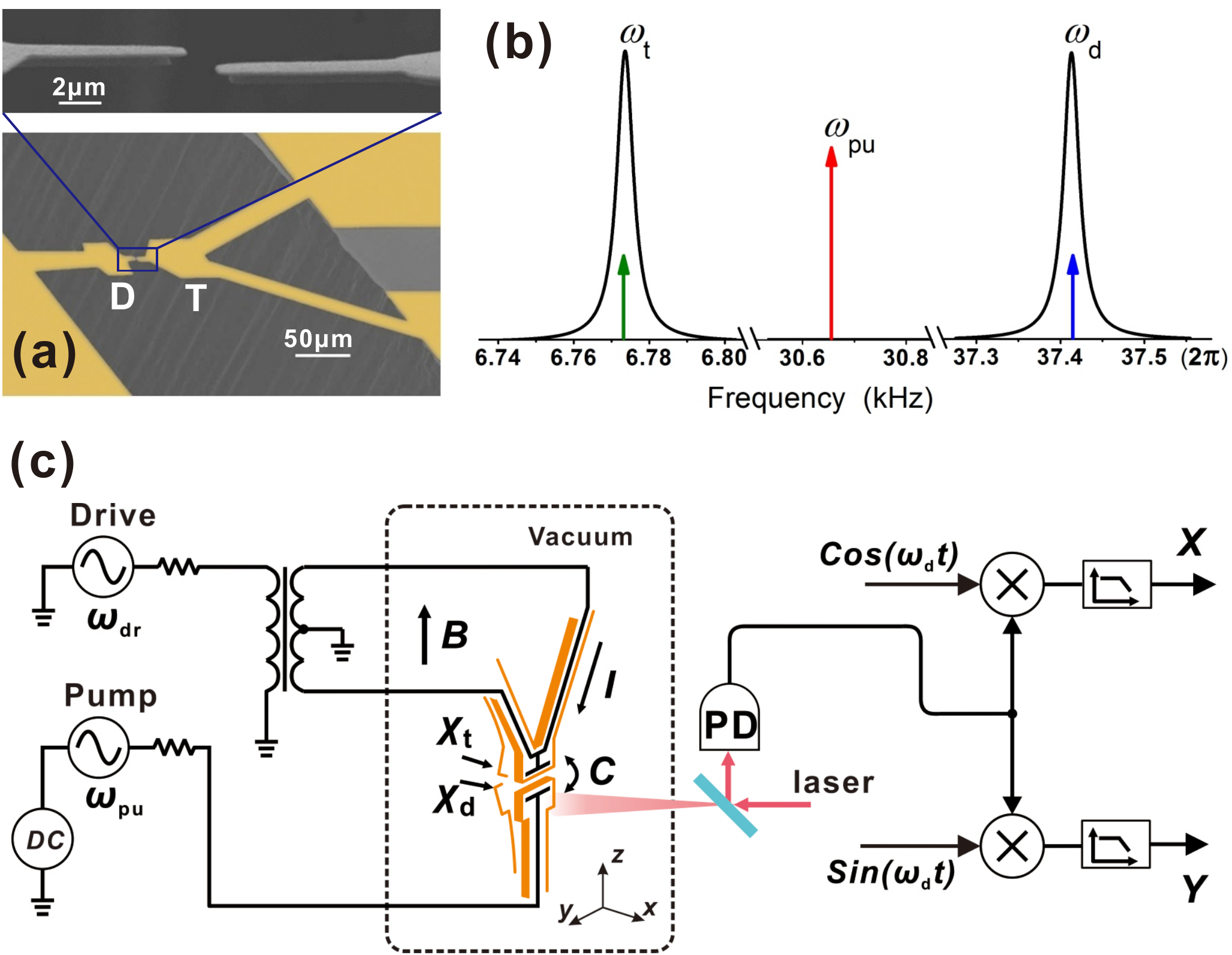}
\caption{(Color online).
(a) Scanning electron microscopy of the coupled resonator system which consists of one single-leg cantilever D
which works as a transducer and a double-leg cantilever T whose motion is under transducing,
with gold on the surface being colored.
(b) A schematic diagram illustrating the thermal noise spectra (green arrow: drive, blue arrow: output of FC) and the frequency of pumping field (red arrow).
(c) A simplified schematic diagram consisting of the circuit of signal generating for the pumping and drive signals (left),
the electric and mechanical operation of the device (middle), and the data processing (right).
A magnetic field $B$ is applied along $z$ axis in order to produce the Lorentz force to drive the resonant T when a voltage of frequency $\omega_{\rm t}$ is between the two legs of resonator T.
A pumping voltage consisting of the superposition of d.c\ and a.c.\ signals is used to generate the coupling between the two resonators via electrostatic force. The effective capacity is of the order of $1$ fF, and the frequency of the a.c. pumping field $\omega_{\rm pu}$ is set to satisfy the relation $\omega_{\rm pu}= \omega_{\rm d} - \omega_{\rm t}$. The reference frequency of demodulation is set at $\omega_{\rm d}$.
}
 \label{fig1}
\end{figure}

Our device to demonstrate the motion transduction and the measurement scheme is illustrated in Fig.\ 1. The device consists of
two silicon mechanical resonators of thickness 1 $\mu$m (Fig.\ 1(a)) which were made of SOI wafer
by double-sided lithography and reactive ion etching; a $100$ nm/$20$ nm of Au/Cr with Cr as
an adhesion layer was evaporated onto the surface to generate a conductive path.
The separation between the two resonator tips was about 1 $\mu$m. Resonator T has two legs connected to external electrodes for the purpose of excitation of motion, so that in the presence of a static magnetic field, the Lorentz force on the small current generated by an a.c.\ drive voltage $V_{\rm dr}\cos(\omega_{\rm dr} t)$ can excite the motion of Resonator T without disturbing Resonator D. By applying a time-dependent pumping voltage $V_{\rm pu}(t)$ between the two resonators, there is an electrostatic force produced, which results in a effective pumping field $\eta(t)=(1/2)(\partial ^{2}C/\partial ^{2}x) V_{\rm pu}(t)^{2}$ in unit of N/m \cite{Rugar2}. Here $C(x_{\rm t}- x_{\rm d})$ is the effective capacity between the two resonators. In practice, $V_{\rm pu}(t)$ is realized by a superposition of a d.c.\ voltage $V_{\rm pu}^{\rm d.c.}$ and an a.c.\ voltage $V_{\rm pu}^{\rm a.c.}\cos(\omega_{\rm pu}t)$ of frequency $\omega_{\rm pu}$ satisfying Eq.\ (\ref{resc}). In this case, only the cross term of $V_{\rm pu}(t)^2$ actually contributes to the FC process due to frequency constrains and the effective coupling strength is then $\eta = 1/2 (\partial ^{2}C/\partial ^{2}x) V_{\rm pu}^{\rm d.c.} V_{\rm pu}^{\rm a.c.}$ (see Supplementary information for details).

The experiments were carried out at room temperature on the device suspended
with springs  in a vacuum chamber of pressure lower than $10^{-8}$ mbar. The basic parameters
of resonators were measured by fiber interferometer which consists of a 1550 nm-distributed-feedback laser diode
using $90/10$ optical coupler, the laser was focused on the end of the resonator by a lens so that resonators T and D
could be measured individually. The resonators T and D have natural resonance frequencies $\omega_{\rm t}= 6.76~$kHz
and $\omega_{\rm d}=$37.41 kHz in absence of externally applied d.c.\ voltage, and quality factors  $Q_{\rm t} = 1200$ and $Q_{\rm d} = 1700$
obtained by free ring down. The thermal noise spectral densities $S_{x,\{\rm t,d\}}^{\rm th}(\omega)$ were measured optically. Then by applying the equipartition theorem \cite{Sidles2} $k_{B}T/k_{\rm t,d}=2\int _{0}^{\infty}S_{x,\{\rm t,d\}}^{\rm th}(\omega)d\omega /2\pi$, the effective spring constants were calculated to be $k_{\rm t}=0.071~$N/m and $k_{\rm d}=1.0~$N/m.  The equation of motion for a damped harmonic oscillator leads to $S_{x,\{\rm t,d\}}^{\rm th}(\omega)=S_{F,\{\rm t,d\}}^{\rm th}\omega_{\rm t,d}^{4}/k_{\rm t,d}^{2}[(\omega_{\rm t,d}^{2}-\omega^{2})^{2}+\omega_{\rm t,d}^2\omega^{2}/Q_{\rm t,d}^{2}]$, from which the frequency-independent thermal force noise density $S_{F,\{\rm t,d\}}^{\rm th}$ can be calculated. From the measured $S_{x,\{\rm t,d\}}^{\rm th}(\omega)$, we obtained $S_{F,\rm t}^{\rm th}= 3.4\times10^{-15}$ N/$\sqrt {\rm Hz}$ and $S_{F,\rm d}^{\rm th} =4.6\times10^{-15}$ N/$\sqrt {\rm Hz}$, and peak displacement noises $5.6\times10^{-11}~$m$/\sqrt {\rm Hz}$ for resonator T and $0.74\times10^{-11}~$m$/\sqrt {\rm Hz}$ for resonator D.

\begin{figure}[hbtp]
\centering
\includegraphics[width=1\columnwidth]{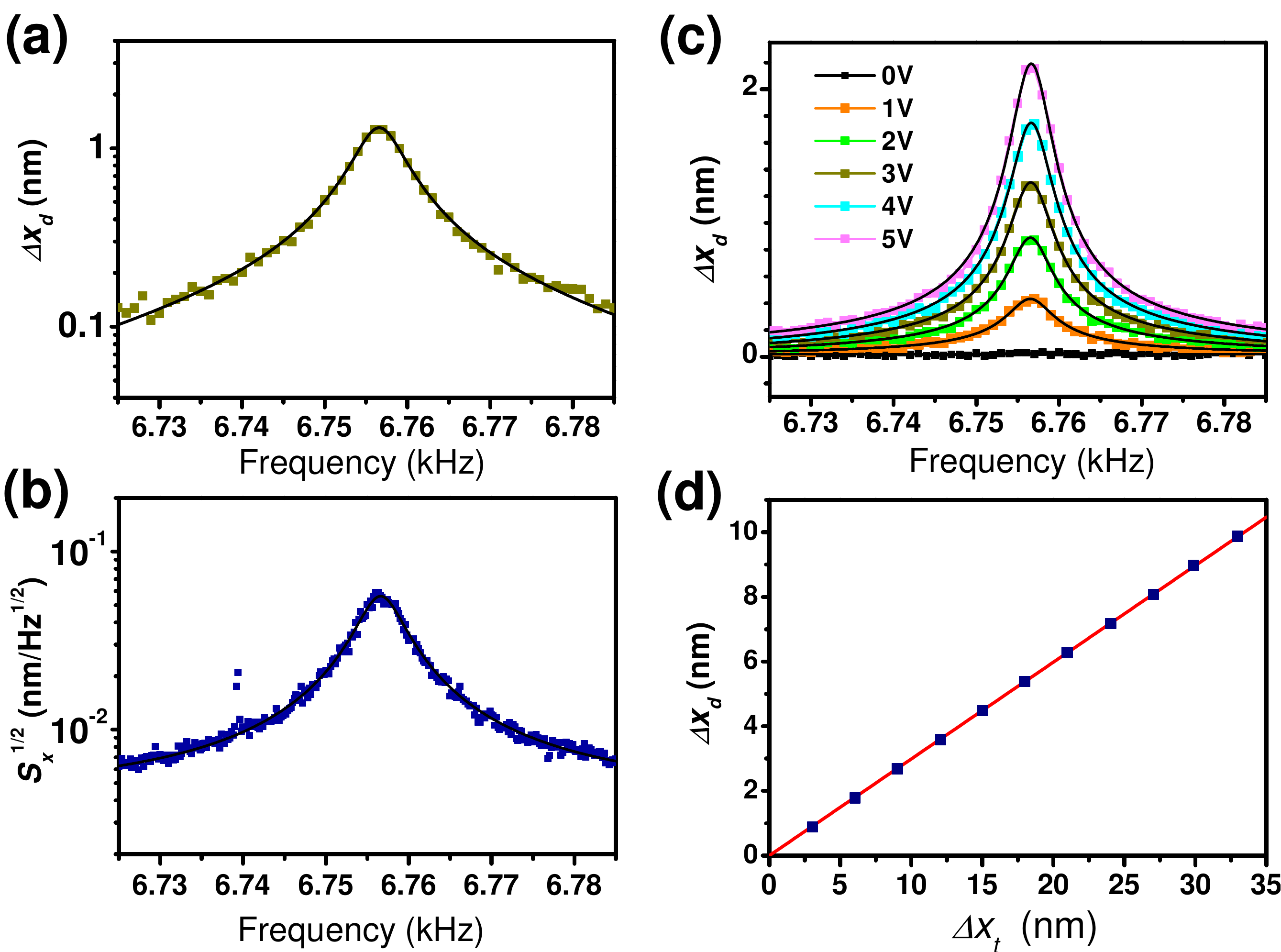}
\caption{(Color online).
(a) Dependence of $\Delta x_{\rm d}$ on the frequency $\omega_{\rm dr}$ of the drive field exerted on resonator T. In the measurement, the  frequency $\omega_{\rm pu}$ was tuned to $\omega_{\rm pu} =\omega_{\rm d} - \omega_{\rm dr}$ when $\omega_{\rm dr}$ was swept.
(b) The optically measured thermal noise spectrum of resonator T.
(c) Same as (a) but for a variety of a.c. pumping voltages $V_{\rm pu}^{\rm a.c.}$. It is noted that for the case $V_{\rm pu}^{\rm a.c.} = 0$, i.e., the a.c.\ pumping voltage being tuned off, $\Delta x_{\rm d}$ is flat at zero, as expected.
(d) The displacements $\Delta x_d$ for resonator D versus $\Delta x_t$  for resonator T under a fixed pumping voltage, with red line being a linear fitting.
The black curves in the sub-figures (a-c) are fits of Lorentz shape to guide eyes.
}
 \label{fig2}
\end{figure}

We first studied the frequency-domain behavior of resonator T under
external excitation by monitoring the motion of resonator
D. In order to do this, the reference frequency in demodulation (Fig.\ 1(c))
was fixed at $\omega_{\rm d}$, and the frequencies of both the
drive and the a.c.\ pump voltages were swept under the constraint
$\omega_{\rm pu}= \omega_{\rm d} - \omega_{\rm dr}$, and the amplitude of $V_{\rm dr}$, $V_{\rm pu}^{\rm d.c.}$
and $V_{\rm pu}^{\rm a.c.}$ were set at $100~\mu$V, $20~$V and $3~$V, respectively.
The amplitude $\Delta x_{\rm d}$ of resonator D was measured optically and plotted
in Fig.\ 2(a) as a function of $\omega_{\rm dr}$. It is important to ascertain that the $\omega_{\rm dr}$ dependence of
$\Delta x_{\rm d}$ actually gives the frequency-domain behavior
 of resonator T. This is confirmed by the fact that the $\Delta x_{\rm d}$ curve in Fig.\ 2(a)
  coincides in both peak frequency and width with the thermal noise spectrum of resonator
   T plotted in Fig.\ 2(b), which was obtained from direct optical measurement.
Measured results of $\Delta x_{\rm d}$ for a series of applied a.c.\ pumping voltages are plotted in Fig.\ 2(c). The measured peak amplitude increases with the increase of $V_{\rm pu}^{\rm a.c.}$ proportionally, indicating that stronger pumping field can be used to produce stronger signal at resonator D for detection. However, deviation from this proportionality becomes apparent for strong enough pumping, as shown later in Fig.\ 3(a). Under a fixed amplitude of
$V_{\rm pu}^{\rm d.c.}$ and $V_{\rm pu}^{\rm a.c.}$ while varying the amplitude of $V_{\rm dr}$, the displacement $\Delta x_{\rm d}$ of resonator D is shown in Fig.\ 2(d)
to be proportional to the displacement $\Delta x_{\rm t}$ of the target resonator T, as expected. These results show that the motion of resonator T is transduced to resonator D via the parametrical coupling process.

\begin{figure}[hbtp]
\centering
\includegraphics[width=1\columnwidth]{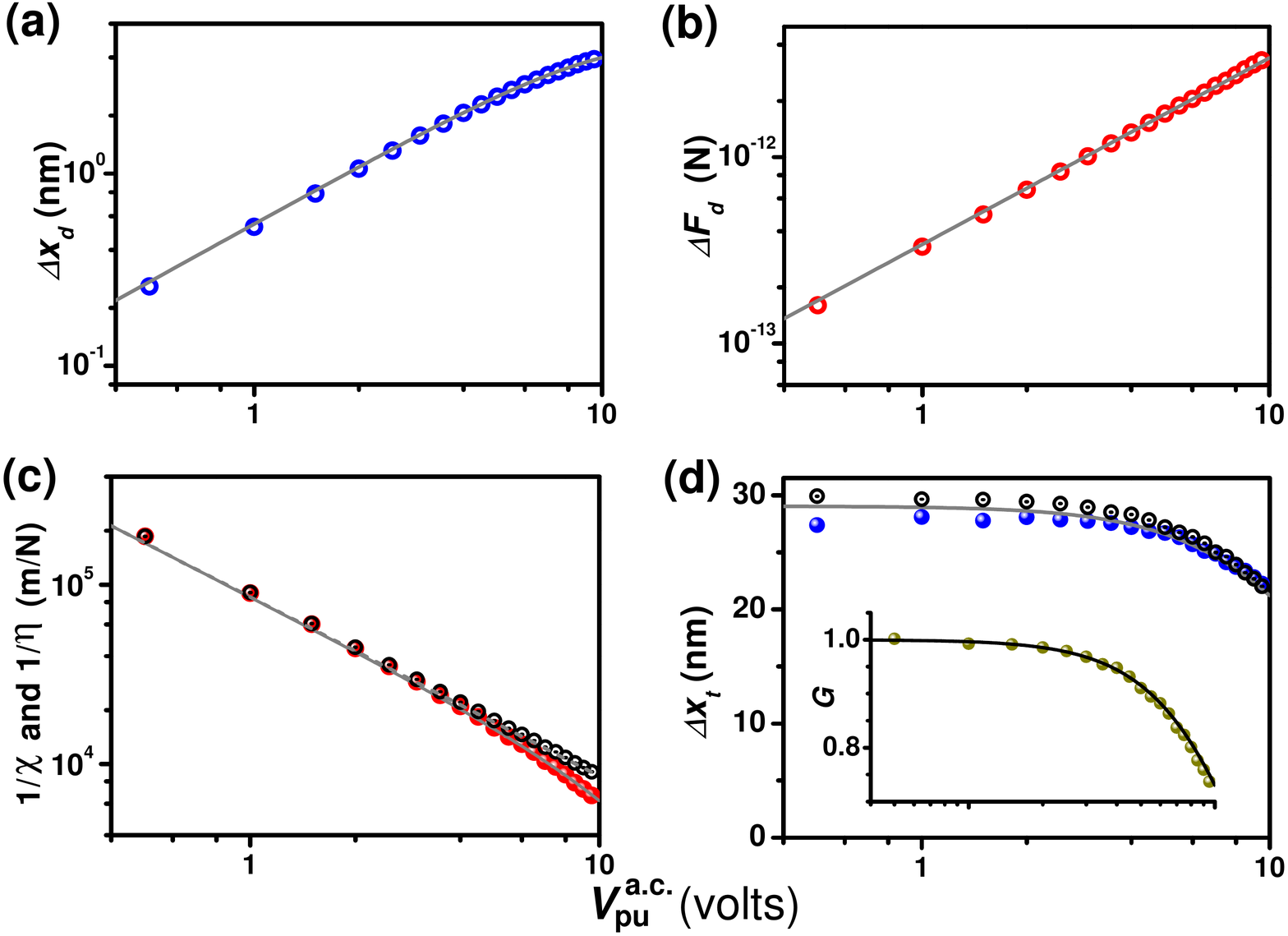}
\caption{(Color online).
(a) Variation of the measured amplitude of resonator D (circles) and the corresponding theoretical curve with a.c.\ pumping voltage.
(b) Variation of the measured force (circles) and the corresponding theoretical value (grey curve) with a.c.\ pumping voltage.
(c) The measured coupling strength $\eta$ (empty circles) and effective transduction factor $\chi$ (solid circles) for different a.c.\ pumping voltages and their theoretical curves.
(d) The calculated motion amplitude of resonator T from transduction process (solid blue circles) and the one directly measured
by optical method (empty black circles) for different a.c.\ pumping voltages and the theoretical value (grey curves), subset is the measured
gain factor $G$ (solid circles) and the theoretical curve.
  \label{fig3}
}
\end{figure}

We further studied the dependence of the transduction process on the coupling strength when resonator T was under coherent resonant excitation and the FC was satisfied, i.e, $\omega_{\rm dr} =\omega_{\rm t}$ and $\omega_{\rm pu} = \omega_{\rm d} - \omega_{\rm t}$. The amplitudes of $V_{\rm dr}$ and $V_{\rm pu}^{\rm d.c.}$ were set
at 100 $\mu$V and 36 V, respectively. The amplitude $V_{\rm pu}^{\rm a.c.}$ was varied from 0.5 to 10 V. The results are plotted in Fig.\ 3.
The displacement response of resonator D in Fig.\ 3(a) shows that under weak pumping the response increases linearly with a.c.\ pumping voltage,
but as the pumping field increases further, nonlinear behavior appears, which fits well to the
theoretical model and gives ${\partial} ^{2}C /{\partial} ^{2}x = 6.5\times 10^{-7}$ F/m$^{2}$.
Fig.\ 3(b) plots the corresponding force felt by resonator D which was calculated by using
$\Delta F_{\rm d} = \Delta x_{\rm d}k_{\rm d}/(G \cdot Q_{\rm d})$, where the gain factor
$G = \Delta x_{\rm t,d}^{\rm pump~on}/\Delta x_{\rm t,d}^{\rm pump~off}$ was obtained from
 the optically measured displacements $\Delta x_{\rm t}^{\rm pump~on}$ and
 $\Delta x_{\rm t}^{\rm pump~off}$. It shows the proportionality of
 $\Delta F_{\rm d}$ to the a.c.\ pumping voltage, as expected.
Fig.\ 3(c) plots the coupling strength $\eta$ derived from measured data by using $\Delta F_{\rm d}/(G \cdot \Delta x_{\rm t})$ and
the effective transduction factor $\chi$ by using $\Delta F_{\rm d}/\Delta x_{\rm t}$. It shows that $1/\chi$ equals $1/\eta$ under
 weak pumping, but is smaller than $1/\eta$  for strong pumping, as a result of the nonlinearity of the transduction process
 (see Supplementary information for details). The motion of resonator T measured by the transduction is compared to the
 one measured directly by optical method in Fig.\ 3(d). The agreement is reasonable well and becomes better when the
 pumping for parametric coupling becomes gradually stronger. The decrease of $\Delta x_{\rm t}$ with the increase of pumping voltage
 clearly shows the decrease in the target resonator's susceptibility to external driving force, by the extent described by the gain factor G plotted as subset of Fig.\ 3(d).

\begin{figure}[htbp]
\centering
\includegraphics[width=.95\columnwidth]{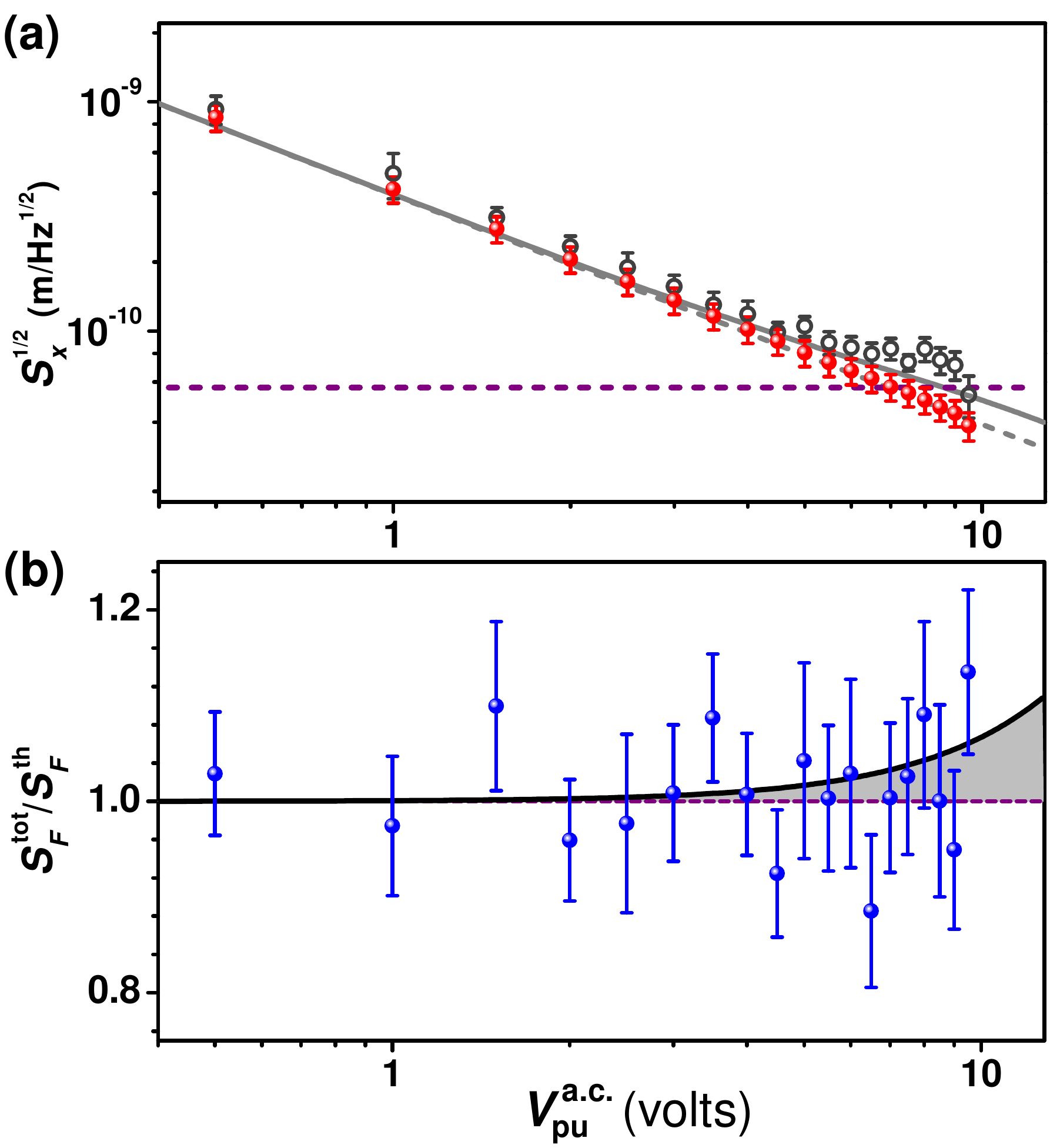}
\caption{(Color online).
(a) Measured noise of the transduction process at a measurement bandwidth of $1$ Hz: imprecision noise (solid circles),
total noise (empty circles), and their corresponding theoretical curves. Error bars show the corresponding statistical fluctuation.
Dashed horizontal line is the thermal noise of resonator T.
(b) The ratio of the measured force noise density $S_{F,\rm d}^{\rm tot}$ under pumping against the intrinsic force
 noise density $S_{F,\rm d}^{\rm th}$ without pumping. The grey regime shows the calculated contributions of the back-action.
  \label{fig4}
}
\end{figure}

Having demonstrated the transduction of motion from resonators T to D, we investigated
the noise of the transduction process. The amplitudes of $V_{\rm dr}$
and $V_{\rm pu}^{\rm d.c.}$ were set the same as those used to obtain Fig.\ 3.
We first studied the transduction noise measured by resonator D.
Fig.\ 4(a) shows the transduction imprecision derived from the optically measured data
 of $S_{F,\rm d}^{\rm tot}$, i.e., the force noise density of resonator D under pumping,
 using the definition $\sqrt{S_x^{\rm im}} = \sqrt{S^{\rm th}_{F,\rm d}}/\eta$ and
the total transduction noise $\sqrt{S_{x}^{\rm tot}} = \sqrt{S_{F,\rm d}^{\rm tot}}/ \chi$.
It shows that when $V_{\rm pu}^{\rm a.c.}=9.5~$V, the imprecision
is $3.9\times10^{-11}$m/$\sqrt{\rm Hz}$, which is smaller than the intrinsic displacements noise of resonator T.
The back-action of the transduction process was studied by measuring optically the force noises of resonator T when the
pumping was off ($\sqrt{S_{F,\rm t}^{\rm th}}$) and on ($\sqrt{S_{F,\rm t}^{\rm tot}}$). The ratio
${S_{F,\rm t}^{\rm tot}}/{S_{F,\rm t}^{\rm th}}$ is plotted in Fig.\ 4(b) together with
theoretical curves. The back-action force noise calculated using $\sqrt {S_{F,\rm t}^{\rm tot}-S_{F,\rm t}^{\rm th}}$
for the measured data when $V_{\rm pu}^{\rm a.c.} = 9.5~$V is only $0.8\times10^{-15}~$N$/\sqrt{\rm Hz}$,
 with a statistic error of $1.4\times10^{-15}~$N$/\sqrt{\rm Hz}$.

In conclusion, we have explored a method to realize motion transduction by the FC process
using parametric coupling between two separate mechanical resonators, and then designed experiments to test
the scheme. Both the transduction imprecision and the back-action
force are substantially smaller than the intrinsic noise level of the resonator being transferred,
and this demonstrates the efficiency of the motion transduction.

Now we discuss the application of the scheme to weak force sensing. Supposing that the motion to be detected is from an ultra-sensitive cantilever (resonator T) with frequency $10~$kHz, force sensitivity the order of $10^{-21}$ N$/\sqrt{\rm Hz}$ and thermal displacement noise the order of $10^{-9}$ m $/\sqrt{\rm Hz}$, and the resonator with frequency about
$1~$MHz and being monitored by a superconducting cavity at about 100 mK \cite{Teufel} can be taken as resonator D, we can realize the motion transduction with back-action force
much smaller than the force noise level of resonator T, while at the same time the motion transduction efficiency is maintained (see Supplementary information for more details).
Combining the transduction scheme with feedback control on resonator D can further improve the transduction sensitivity and realize dynamical cooling of resonator T, which is out of the scope of this work.

We thank Fei Xue for help in setting up the experimental hardwares and Yiqun Wang from Suzhou Institute of Nano-Tech and Nano-Bionics for fabricating the mechanical resonators.
This work was supported by the National Key Basic Research Program of China (Grant No. 2013CB921800), the National Natural Science Foundation of China (Grant Nos. 11227901, 91021005, 11274299, 11104262 and 10834005) and the `Strategic Priority Research Program (B)' of the CAS (Grant No. XDB01030400).

\end{document}